\documentclass{iau} 
\usepackage{graphicx}
\usepackage{natbib}
\usepackage{url}
\usepackage{hyperref}
\hypersetup{
    colorlinks=true,
    linkcolor=blue,
    filecolor=magenta,      
    urlcolor=magenta,
    citecolor=blue,
    }
\usepackage[usenames, dvipsnames]{color, xcolor}

\pubyear{2022}
\volume{370}
\jname{Winds of Stars and Exoplanets}
\editors{A.~A. Vidotto, L.~Fossati \& J.~Vink, eds.}

\title[A new grid of magnetic stellar evolution models] 
{Spin-down and reduced mass loss in early-type stars with large-scale magnetic fields}
\author[Z. Keszthelyi et al.]   
{Z. Keszthelyi$^{1,2}$,
A. de Koter$^{1,3}$,
Y. G\"otberg$^{4}$, 
G. Meynet$^{5}$,
S.A.~Brands$^{1}$,
V. Petit$^{6}$,  
M. Carrington$^{7}$, 
A. David-Uraz$^{8,9}$, 
S.T.~Geen$^{1}$, 
C. Georgy$^{5}$,
R. Hirschi$^{10,11}$, 
J. Puls$^{12}$,  
K.J.~Ramalatswa$^{13,14}$,
M.E. Shultz$^{6}$, 
A. ud-Doula$^{15}$}
\affiliation{$^{1}$Anton Pannekoek Institute for Astronomy, University of Amsterdam, Science Park 904, 1098~XH, Amsterdam, The Netherlands \\ [\affilskip]
$^{2}$Center for Computational Astrophysics, Division of Science, National Astronomical Observatory of Japan, 2-21-1, Osawa, Mitaka, Tokyo 181-8588, Japan \\ [\affilskip]
$^{3}$Institute of Astronomy, KU Leuven, Celestijnenlaan 200D, 3001 Leuven, Belgium \\ [\affilskip]
$^{4}$The observatories of the Carnegie institution for science, 813 Santa Barbara Street, Pasadena, CA 91101, USA \\ [\affilskip]
$^{5}$Geneva Observatory, University of Geneva, Maillettes 51, 1290 Sauverny, Switzerland \\ [\affilskip]
$^{6}$Dept. of Physics and Astronomy, Bartol Research Institute, University of Delaware, 217 Sharp Lab, Newark, DE 19716, USA \\ [\affilskip]
$^{7}$Dept. of Physics and Space Science, Royal Military College of Canada, PO Box 1700, Station Forces, Kingston, ON K7K 0C6, Canada \\  [\affilskip]
$^{8}$Dept. of Physics and Astronomy, Howard University, Washington, DC 20059, USA \\ [\affilskip]
$^{9}$Center for Research and Exploration in Space Science and Technology, and X-ray Astrophysics Laboratory, NASA/GSFC, Greenbelt, MD 20771, USA \\ [\affilskip]
$^{10}$Astrophysics Group, Keele University, Keele, Staffordshire ST5 5BG, UK\\ [\affilskip]
$^{11}$Institute for Physics and Mathematics of the Universe (WPI), University of Tokyo, 5-1-5 Kashiwanoha, Kashiwa 277-8583, Japan \\ [\affilskip]
$^{12}$LMU M\"unchen, Universit\"atssternwarte, Scheinerstr. 1, 81679 M\"unchen, Germany \\  [\affilskip]
$^{13}$Dept. of Astronomy, University of Cape Town, Private Bag X3, Rondebosch 7701, South~Africa \\ [\affilskip]
$^{14}$South African Astronomical Observatory, PO Box 9, Observatory, 7935, South~Africa \\ [\affilskip]
$^{15}$Dept. of Physics, Penn State Scranton, 120 Ridge View Drive, Dunmore, PA 18512, USA \\ }

\begin{document}

\maketitle

\begin{abstract}
Magnetism can greatly impact the evolution of stars. In some stars with OBA spectral types there is direct evidence via the Zeeman effect for stable, large-scale magnetospheres, which lead to the spin-down of the stellar surface and reduced mass loss. So far, a comprehensive grid of stellar structure and evolution models accounting for these effects was lacking. For this reason, we computed and studied models with two magnetic braking and two chemical mixing schemes in three metallicity environments with the \textsc{mesa} software instrument. We find notable differences between the subgrids, which affects the model predictions and thus the detailed characterisation of stars. We are able to quantify the impact of magnetic fields in terms of preventing quasi-chemically homogeneous evolution and producing slowly-rotating, nitrogen-enriched (``Group~2'') stars. Our model grid is fully open access and open source.  
\keywords{stars: evolution --- stars: massive ---
stars: magnetic field --- stars: rotation --- stars: abundances}
\end{abstract}

\firstsection 
              
\section{Introduction}

Magnetism is a key component in several astrophysical phenomena. For example, magnetic fields play a crucial role in regulating star formation and controlling the formation of neutron stars (e.g., \citealp{C11,T11}). A fraction of massive stars (initially $M_\star > 8$~M$_\odot$) and intermediate-mass stars (initially 3~M$_\odot < M_\star < 8$~M$_\odot$) show evidence of stable, globally organised, large-scale magnetic fields that are understood to be of fossil origin. The exact origin of such fossil fields is unclear; however, they may result from the pre-main sequence evolution of the star or from stellar merger events \citep{S19}.
Fossil fields form a magnetosphere around the star, affecting the wind and stellar rotation. For their stability, they must be anchored in deep stellar layers.
While it is known that magnetic fields can significantly impact the physics and evolution of stars, there has been no comprehensive grid of stellar evolution models of massive stars that would take into account the effects of surface fossil magnetic fields.
We use the \textsc{mesa} software instrument \citep{P19} to compute models and map out a large parameter space. The main new additions in our calculations are magnetic mass-loss quenching, magnetic braking, and efficient angular momentum transport (see next section).

\section{Background}

\subsection{Alfv\'en radius}\label{sec:alf}

The Alfv\'en radius characterises a critical distance at which the magnetic energy density and the gas kinetic energy density are equal.  \cite{ud09} use a numerical fitting to quantify the Alfv\'en radius as: 
\begin{equation}\label{eq:alf1}
    \frac{R_{\rm A}}{R_\star} \approx 1 + (\eta_\star + 0.25)^{x} - (0.25)^{x} \, ,
\end{equation}
with $x=1/4$ and $1/6$ for dipolar and quadrupolar field geometries, which we assume in our INT and SURF models (see next section), respectively. $R_\star$ is the stellar radius.
The equatorial magnetic confinement parameter $\eta_\star$ is defined as: 
\begin{equation}
    \eta_\star = \frac{B_{\rm eq}^2 R_\star^2}{\dot{M}_{B=0} \cdot v_\infty} \, ,
\end{equation} 
\noindent where $B_{\rm eq}$ is the equatorial magnetic field strength, $\dot{M}_{B=0}$ is the mass-loss rate in absence of a magnetic field, and $v_\infty$ is the terminal velocity \citep[][]{ud09}.

\subsection{Mass-loss quenching}

Large-scale magnetic fields lead to channelling and trapping the wind plasma within the magnetosphere. 
To account for the global, time-averaged effect of this process, the adopted mass-loss rates in our models are reduced by a parameter $f_{\rm B}$, which is defined following the works of \cite{ud08,ud09}:
\begin{equation}\label{eq:fb1}
f_{\rm B} = \, \frac{\dot{M}}{\dot{M}_{B=0}} = \, 1 - \sqrt{1 - \frac{1}{R_{\rm c}}}  \quad \mathrm{if} \quad R_{\rm A} < R_{\rm K}
\end{equation}
and  
\begin{equation}\label{eq:fb2}
f_{\rm B} = \, \frac{\dot{M}}{\dot{M}_{B=0}} \, = 2 - \sqrt{1 - \frac{1}{R_{\rm c}}} - \sqrt{1 - \frac{0.5}{R_{\rm K}}}    \quad \mathrm{if} \quad R_{\rm K}  < R_{\rm A} 
\end{equation}
\noindent where $R_{\rm A} $, $R_{\rm K} $, and $R_{\rm c} $ are the Alfv\'en radius, the Kepler co-rotation radius, and the closure radius in units of the stellar radius, respectively (see \citealp{K19,K20,K21}, and references therein). 

\subsection{Magnetic braking}

Stellar rotation exerts a force on magnetic field lines and causes the field to bend in the azimuthal direction. The associated Maxwell stresses are very efficient at transferring angular momentum to the surrounding wind plasma, which results in slowing the spin of the star.
We quantify this process following the work of \cite{ud09}, where the total wind and magnetic field-induced loss of angular momentum can be expressed via a Weber-Davis \citep{wd67} scaling relation:
\begin{equation}\label{eq:br}
\frac{\mathrm{d}J_{\rm B}}{\mathrm{d}t}  = \frac{2}{3} \dot{M}_{B=0} \,  \Omega_\star R_{A}^2 \, ,
\end{equation}
\noindent with $\mathrm{d}J_{\rm B}/\mathrm{d}t$ the rate of angular momentum loss from the system, $\Omega_\star$ the surface angular velocity, and $R_{A}$ the Alfv\'en radius (defined in Equation~\ref{eq:alf1}).

Based on how magnetic braking is applied, we split the model calculations into two branches. In one case, we assume internal magnetic braking (INT models). These models are solid-body rotating and specific angular momentum is extracted from each layer of the stellar model.
In the other case, we only allow the model to directly remove specific angular momentum from the upper envelope of the star (SURF models). Thus, radial differential rotation can develop in deeper layers in the SURF models. 

\subsection{Chemical mixing}

Due to stellar rotation, chemical elements are also mixed in radiative regions of stars. Consequently, main sequence massive stars can replenish their core with more hydrogen, and also enrich their surface with core-produced materials (most importantly, nitrogen). We adopt a diffusive scheme, following the work of \cite{P89}, to account for these effects. 
\begin{equation}\label{eq:diff}
 \frac{ \partial X_i }{ \partial t}    = \frac{\partial}{\partial m} \left[ (4 \pi r^2 \rho)^2  \, D_{\rm chem} \,  \frac{\partial X}{\partial m}   \right] + \left( \frac{\mathrm{d} X_i}{\mathrm{d} t} \right)_{\rm nuc}  \, , 
\end{equation}
\noindent where $X_i$ is the mass fraction of a given element $i$, $t$ is the time, $m$ and $\rho$ are the mass coordinate and mean density at a given radius $r$, $D_{\rm chem}$ is the sum of individual diffusion coefficients contributing to chemical mixing, and the final term accounts for nuclear burning. 

The detailed mixing processes remain highly uncertain in massive star evolutionary models. For this reason, we also split the model calculations to construct $D_{\rm chem}$ in two different ways.
In one case (Mix1), we assume chemical mixing equations that are commonly adopted in \textsc{mesa}. We thus adopt $D_{\rm chem}$ as a sum of individual diffusion coefficients describing meridional circulation, shear, and GSF instabilities. We then scale $D_{\rm chem}$ with commonly used factors in this approach.
In the other case (Mix2), we implement the mixing equations of \cite{Z92} into the \textsc{mesa} code. Here we use the vertical shear mixing and meridional circulation (both defined differently than in the above approach) to construct $D_{\rm chem}$. In this case, scaling factors are not used. We do not change the equations for angular momentum transport for the sake of a consistent model-to-model comparison.

\section{Summary of computed models}

In total, we computed 8,748 main sequence stellar evolution models, with 3 stellar structure models for each of these corresponding to ZAMS, mid-MS, and TAMS evolutionary stages. We also generated isochrones. All the computed models are open access and open source and available on Zenodo at \url{https://doi.org/10.5281/zenodo.7069766}.
The initial mass and equatorial magnetic field strength spans from 3 to 60~M$_\odot$ and from 0 to 50~kG in our grid of models. Four sub-grids are available, accounting for two magnetic braking and two chemical mixing schemes, introduced above.
Hot star mass-loss rates in our models are adopted from \cite{V01} and are reduced by a factor of 2 to account for the general trend  evidenced from recent clumping-corrected mass-loss rate determinations. Mass-loss rates lower than the predictions of \cite{V01} are also expected from new numerical simulations \citep{B20,Kr21}. 
The metallicity in our model grid corresponds to Solar, LMC, and SMC abundances, respectively. 

\section{Results \& Discussion}

\begin{figure}[ht!]
\begin{center}
 \includegraphics[width=0.35\textwidth]{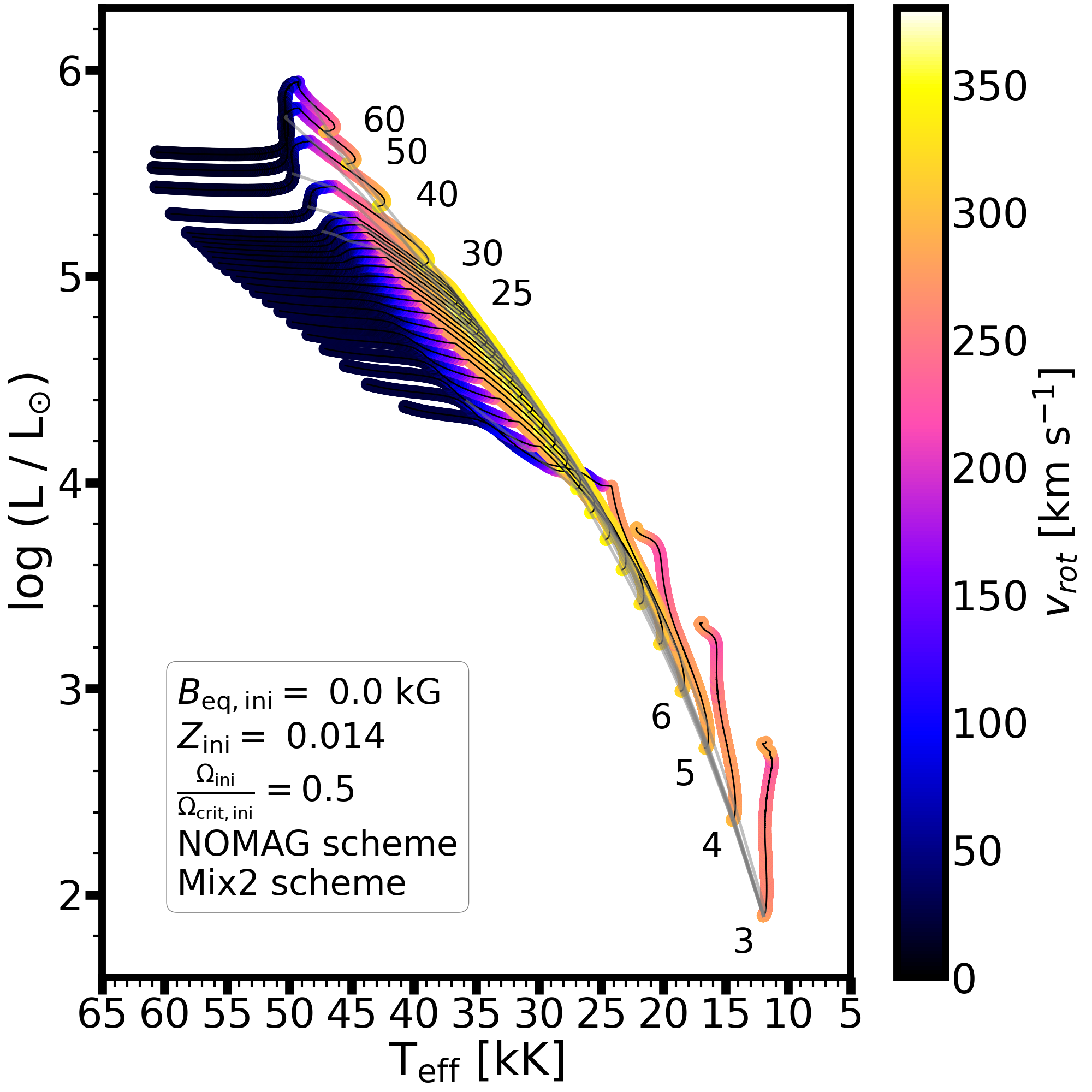}\includegraphics[width=0.35\textwidth]{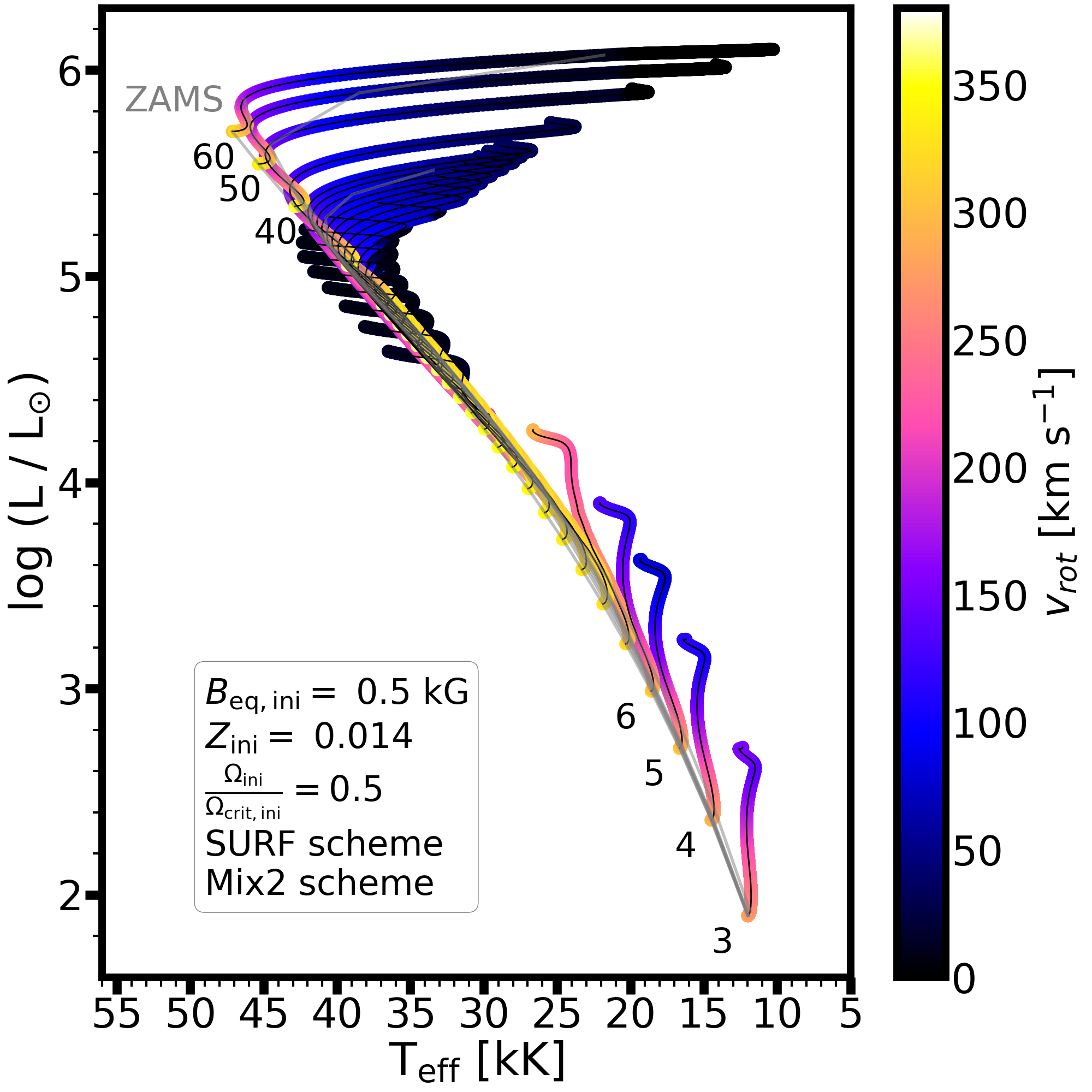}\includegraphics[width=0.35\textwidth]{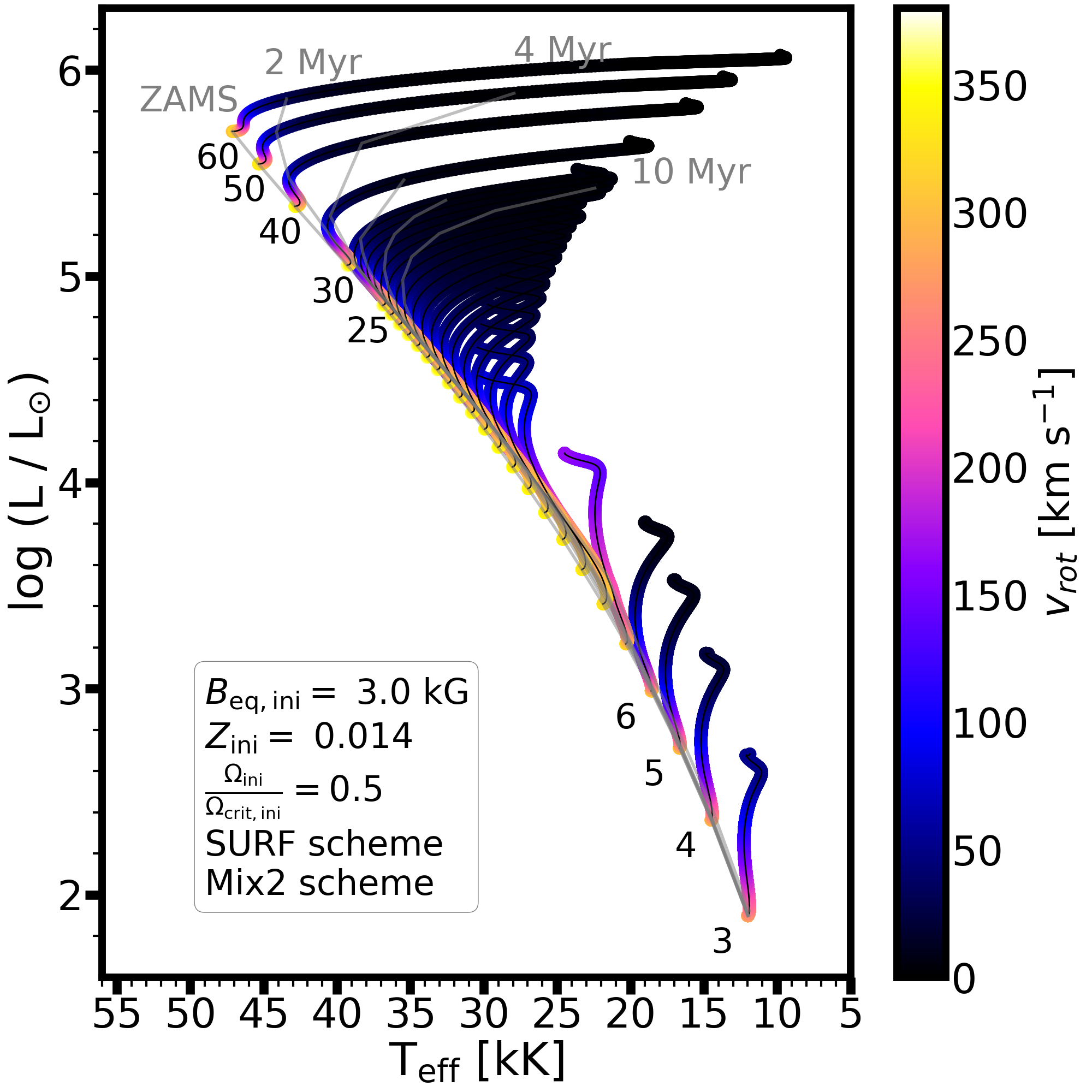}
 \caption{HRDs of the computed models at solar metallicity with an initial ratio of 0.5 critical angular velocity within the efficient Mix2 chemical mixing scheme. Panels from left to right show models with initial equatorial magnetic field strengths of 0, 0.5, 3 kG, respectively. The colour-coding corresponds to the surface rotational velocity. The initial masses in solar units are indicated next to the ZAMS. Between 6 and 25~M$_\odot$, the increment is 1~M$_\odot$. From \cite{K22}.} \label{fig1}
\end{center}
\end{figure}

\begin{figure}[ht!]
\begin{center}
 \includegraphics[width=1\textwidth]{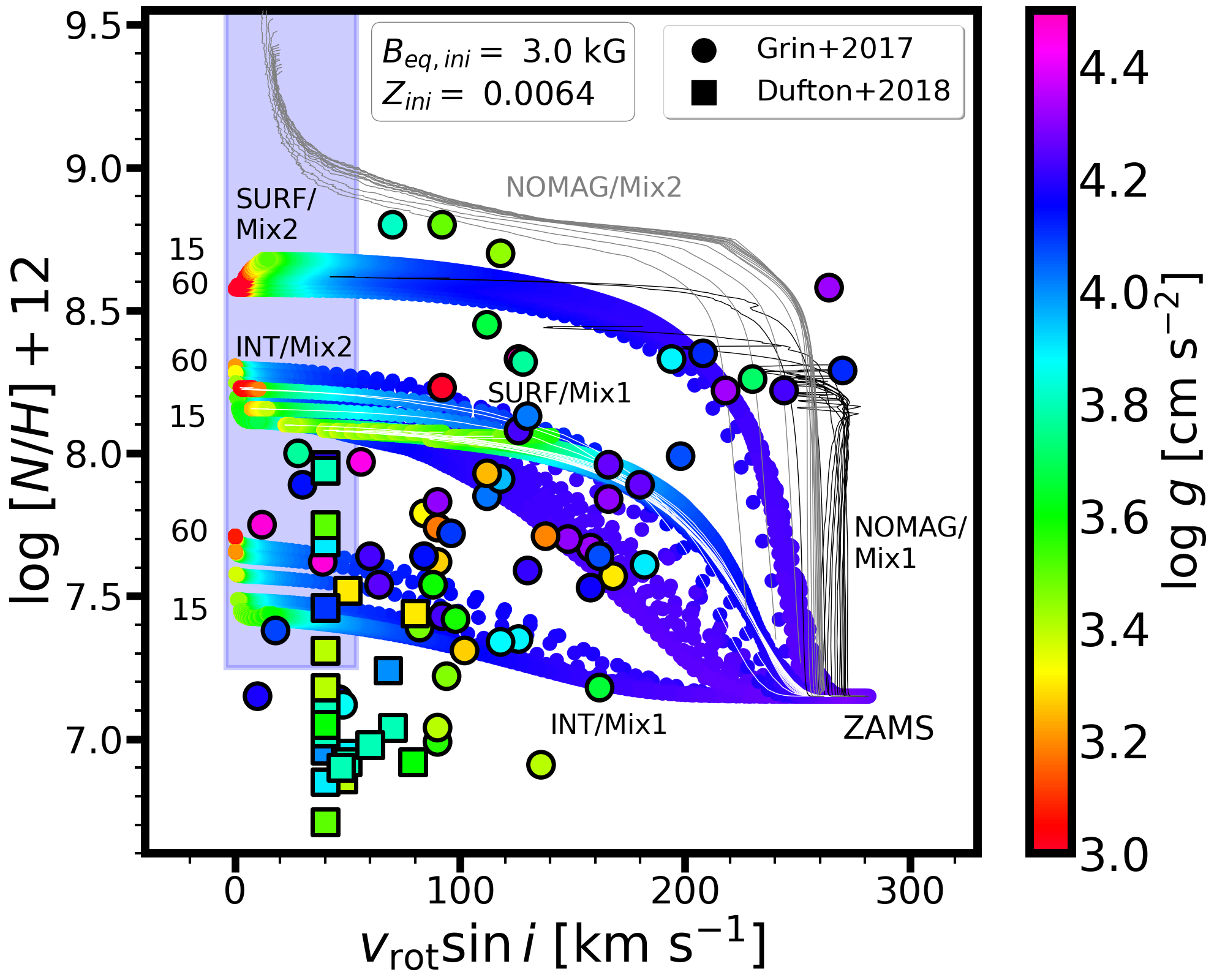}
 \caption{Hunter diagram of magnetic single-star evolutionary models with $B_{\rm eq, ini}= 3$~kG, $\Omega_{\rm ini}/\Omega_{\rm ini, crit}= 0.5$ at LMC ($Z_{\rm ini}= 0.0064$) metallicity within two magnetic braking and two chemical mixing schemes. Models within the SURF/Mix1 scheme are also shown with white lines since they overlap with the INT/Mix2 models. Additionally, two sets of non-magnetic (NOMAG) models are shown within the Mix1 (grey) and Mix2 (black) schemes. Models with initial masses from 15 to 60~M$_\odot$ are shown. The coloured area corresponds to our definition of Group 2 stars ($v \sin i < 50$~km\,s$^{-1}$ and at least 0.1 dex nitrogen enrichment in spectroscopic units). The colour-coding of the models shows the logarithmic surface gravity. Observations are shown with circles and squares. A typical reported uncertainty in the observed nitrogen abundances is about 0.1 dex. From \cite{K22}.} \label{fig2}
\end{center}
\end{figure}

\subsection{Quasi-chemically homogeneous evolution}

When internal chemical mixing is efficient, our models at solar metallicity with an initial rotation rate of 0.5 critical angular velocity show a blueward evolution on the HRD (left panel of Figure~\ref{fig1}). This is because all stellar layers have a nearly homogeneous chemical composition. Given the blueward evolution, the increase in effective temperature will lead to Wolf-Rayet type mass-loss rates in our models, which can help i) spin down the stellar models and ii) decrease the surface hydrogen abundance. 
With an initial equatorial magnetic field strength of 0.5~kG (middle panel), the models experience a short blueward evolution. However, within a spin-down timescale, the decrease of rotational velocity results in less efficient chemical mixing. Therefore, the models turn to a redward evolution on the main sequence.
With $B_{\rm eq, ini} = 3$~kG, the initial blueward evolution is shorter, and most of the main sequence evolution follows a classical path. We can thus conclude that surface fossil magnetic fields may play an important role in preventing quasi-chemically homogeneous main sequence evolution, which may be an important channel for several astrophysical phenomena (e.g., \citealp{Y06}). 

\subsection{Magnetic stellar evolution models on the Hunter diagram}

The Hunter diagram \citep{H08}, showing surface nitrogen abundance as a function of projected rotational velocity, is an important diagnostic tool to gain insights into the chemical and rotational evolution of stars. The efficiency of chemical mixing and magnetic braking play an important role in shaping the quantitative evolution of the models on this diagram.
For LMC metallicity, we assumed a baseline abundance of $\log [N/H] = 7.15$. The uncertainties in the magnetic models may lead to an order of magnitude difference in the predicted nitrogen abundances for a typical 3~kG initial equatorial magnetic field strength (Figure~\ref{fig2}).
Nonetheless, we find that these magnetic models, regardless of the uncertainties related to mixing and braking schemes, produce slowly-rotating, nitrogen-enriched stars. These so-called ``Group 2'' stars could not be explained with typical single star evolution thus far, though they are commonly found in spectroscopic samples.
We note however that spectropolarimetric observations and comprehensive magnetic characterisation of Group 2 stars are largely lacking in the Galaxy (see however, e.g., \citealp{A14,M12,M15}) and are practically unavailable in the Magellanic Clouds. Therefore Group 2 stars might be prime targets to detect magnetic fields.

\section{Conclusions}

We computed and studied an extensive grid of stellar evolution and structure models, incorporating the effects of surface fossil magnetic fields. We can quantitatively demonstrate that i) quasi-chemically homogeneous evolution could be mitigated and prevented for increasing magnetic field strength, and ii) slowly-spinning, nitrogen-enriched Group~2 stars on the Hunter diagram could be produced by magnetic massive stars. The library of stellar models is available to the community via Zenodo at \url{https://doi.org/10.5281/zenodo.7069766}.



\begin{thebibliography}{}

\bibitem[Aerts et al. (2014)]{A14}
{{Aerts}, C., {Molenberghs}, G., {Kenward}, M.~G., {Neiner}, C.}, 2014, \textit{ApJ} 781, 88

\bibitem[Bj{\"o}rklund et al. (2020)]{B20}
{{Bj{\"o}rklund}, R., {Sundqvist}, J.~O., {Puls}, J., {Najarro}, F.}, 2020, \textit{A\&A} 648, A36

\bibitem[Commer{\c{c}}on et al. (2011)]{C11}
{{Commer{\c{c}}on}, B., {Hennebelle}, P., {Henning}, T.}, 2011, \textit{ApJL} 742, L9

\bibitem[Hunter et al. (2008)]{H08}
{{Hunter}, I., {Brott}, I., {Lennon}, D.~J., {Langer}, N., {Dufton}, P.~L., et al.}, 2008, \textit{ApJL} 676, L29

\bibitem[Keszthelyi et al. (2019)]{K19}
{{Keszthelyi}, Z., {Meynet}, G., {Georgy}, C., et al.}, 2019, \textit{MNRAS} 485, 5843

\bibitem[Keszthelyi et al. (2020)]{K20}
{{Keszthelyi}, Z., {Meynet}, G., {Shultz}, M.~E., et al.}, 2020, \textit{MNRAS} 493, 518

\bibitem[Keszthelyi et al. (2021)]{K21}
{{Keszthelyi}, Z., {Meynet}, G., {Martins}, F., et al.}, 2021, \textit{MNRAS} 504, 2474

\bibitem[Keszthelyi et al. (2022)]{K22}
{{Keszthelyi}, Z., {de~Koter}, A., {G\"otberg}, Y., et al.}, 2022, \textit{MNRAS} 517, 2028

\bibitem[Krti{\v{c}}ka et al. (2021)]{Kr21}
{{Krti{\v{c}}ka}, J., {Kub{\'a}t}, J., {Krti{\v{c}}kov{\'a}}, I.}, 2021, \textit{A\&A} 647, A28 

\bibitem[Martins et al. (2015)]{M15}
{{Martins}, F., {Herv{\'e}}, A., {Bouret}, J.-C., {Marcolino}, W., {Wade}, G.~A., et al.}, 2015, \textit{A\&A} 575, A34

\bibitem[Martins et al. (2012)]{M12}
{{Martins}, F., {Escolano}, C., {Wade}, G.~A., {Donati}, J.~F., {Bouret}}, 2012, \textit{A\&A} 538, A29

\bibitem[Paxton et al. (2019)]{P19}
{{Paxton}, B., {Smolec}, R., {Schwab}, J., {Gautschy}, A.,
         {Bildsten}, L., et al.}, 2019, \textit{ApJS} 243, 10

\bibitem[Pinsonneault et al. (1989)]{P89}
{{Pinsonneault}, M.~H., {Kawaler}, S.~D., {Sofia}, S., {Demarque}, P.}, 1989, \textit{ApJ} 338, 424

\bibitem[Schneider et al. (2019)]{S19}
{{Schneider}, F.~R.~N., {Ohlmann}, S.~T., {Podsiadlowski}, P.}, et al., 2019, \textit{Nat} 574, 211

\bibitem[ud-Doula et al. (2008)]{ud08}
{{ud-Doula}, A., {Owocki}, S.~P., {Townsend}, R.~H.~D.}, 2008, \textit{MNRAS} 385, 97

\bibitem[ud-Doula et al. (2009)]{ud09}
{{ud-Doula}, A., {Owocki}, S.~P., {Townsend}, R.~H.~D.}, 2009, \textit{MNRAS} 392, 1022

\bibitem[Takiwaki \& Kotake (2011)]{T11}
{{Takiwaki}, T., {Kotake}, K}, 2011, \textit{ApJ} 743, 30

\bibitem[Vink et al. (2001)]{V01}
{{Vink}, J.~S., {de Koter}, A., {Lamers}, H.~J.~G.~L.~M.}, 2001, \textit{A\&A} 369, 574

\bibitem[Weber \& Davies (1967)]{wd67}
{{Weber}, E.~J., {Davis}, Jr., L.}, 1967, \textit{ApJ} 148, 217

\bibitem[Yoon et al. (2006)]{Y06}
{{Yoon}, S. -C., {Langer}, N., {Norman}, C.}, 2006, \textit{ApJ} 460, 199

\bibitem[Zahn (1992)]{Z92}
{{Zahn}, J.-P.}, 1992, \textit{A\&A} 265, 115

\end{thebibliography}

\end{document}